\documentclass[a4paper,12pt]{article}

\usepackage{amsmath,amssymb}

\def \field {{\cal F}(u,D)}
\def \ring {{\cal R}(u,D)}
\def \sring {\hat{{\cal R}}(u,\eta)}
\def \ord {\, \mbox{ord}}
\def \sign {\, \mbox{sign}}

\newtheorem{Def}{Definition}
\newtheorem{The}{Theorem}
\newtheorem{Pro}{Proposition}
\newtheorem{Lem}{Lemma}

\begin{document}
\small
\title{Perturbative Symmetry Approach.}
\author{A.V. Mikhailov
\thanks{Applied Mathematics Department, University of Leeds,UK and
L.D. Landau Institute for Theoretical Physics, Moscow, Russia.}
, V.S. Novikov\thanks
{L.D. Landau Institute for Theoretical Physics, Moscow, Russia, e-mail:
nvs@itp.ac.ru}}

\maketitle

\section{Introduction}

In the Symmetry Approach the existence of infinite hierarchies of
higher symmetries and/or local conservation laws is taken as a
definition of integrability. The main aims of the approach are to
obtain easily verifiable necessary conditions of integrability,
to identify integrable cases and even to give complete
description and classification of integrable system of a
particular type. It proved to be quite successful for a
description of integrable evolution equations and systems of
equations \cite{ss,msy,mss}. Further progress has been achieved
in \cite{w,sw} where the symbolic representation of the ring of
differential polynomials enables to reduce the problem of
classification of polynomial homogeneous partial differential
equations with higher symmetries to a problem of factorisation of
very special symmetrical polynomials and the latter has been
solved by Number Theory methods. The aim of our paper is to
formulate a perturbative version of the symmetry approach in the
symbolic representation and to generalise it in order to make it
suitable for study nonlocal and non-evolution equations. We apply
our theory to describe integrable generalisations of the
Benjamin-Ono type equations and to isolate integrable cases of
Camassa-Holm type equations.

\section{Symmetry Approach - basic definitions and facts}

 Here we give a brief account of basic
facts, definitions and notations (details and proofs see in
\cite{ss,msy,mss}).

Suppose we have an evolution partial differential equation
\begin{equation}
\label{eq1}
u_t=F(u_{n},...,u_1, u_0)\, ,
\quad n\ge 2
\end{equation}
where $u_0=u(x,t),u_1=u_x (x,t),u_2=u_{xx}(x,t), \ldots ,u_n=
\partial^n_x u(x,t)$. Equation (\ref{eq1}) can be represented by two
compatible infinite dimensional dynamical systems
\begin{eqnarray*}
&&u_{0x}=u_1,\ u_{1x}=u_2,\ \ldots\ ,\ u_{mx}=u_{m+1},\ \ldots\\
&&u_{0t}=F_0,\ u_{1t}=F_1,\ \ldots\ ,\  u_{mt}=F_m,\ \ldots
\end{eqnarray*}
where
\[ F_k (u_{n+k},...,u_1, u_0)=D^k (F (u_{n},...,u_1, u_0)) \]
and the linear differential operator
\begin{equation}\label{D}
D=u_1\frac{\partial}{\partial u_{0}}+ u_2\frac{\partial} {\partial
u_{1}}+ u_3\frac{\partial}{\partial u_{2}}+\cdots\, ,
\end{equation}
represents the derivation in $x$. The operator $D$ is applied to
functions of finite number of variables and therefore only finite
number of terms in the sum (\ref{D}) is required.

In the Symmetry Approach it is assumed that all functions such as
$F_k$ depends on a finite number of variables and belong to a
proper differential field $\field $ generated by $u$ and the
derivation $D$ (\ref{D}). Partial differential equation
(\ref{eq1}) defines another derivation of the field $\field$

\begin{equation}
\label{Dt}
D_t=F\frac{\partial}{\partial u_{0}}+ F_1\frac{\partial} {\partial
u_{1}}+F_2\frac{\partial}{\partial u_{2}}+ \cdots\, , \qquad F_k\in
\field
\end{equation}
commuting with $D$.

A symmetry of equation (\ref{eq1}) can be defined as derivation
$D_\tau$
\begin{equation}
\label{Dtau}
D_\tau=G\frac{\partial}{\partial u_{0}}+ G_1\frac{\partial} {\partial
u_{1}}+ G_2\frac{\partial}{\partial u_{2}}+ \cdots\, , \qquad G_k\in
\field
\end{equation}
of our field $\field$ which commutes with derivations $D$ and $D_t$.
It follows from $[D,D_\tau ]=0$ that $G_k=D^k(G)$.

The Fr\'echet derivative of  $a\in \field$ is defined as a linear
differential operator of the form $$ a_*=\sum_k\frac{\partial
a}{\partial u_k}D^k \, . $$ We say that element $a$ has order $n$
if the corresponding differential operator $a_*$ is of order $n$.
Order of equation (\ref{eq1}) is the order of $F$, the order of a
symmetry is defined in a similar way. If a symmetry has order
$n\ge 2$ then we call it a higher symmetry. The Lie brackets for
any two elements $a,b\in \field$ is defined as
\[
[a,b]=a_*(b)-b_*(a) \, .
\]
In these terms the definition of symmetry of equation (\ref{eq1}) can
be formulated as follows: function $G\in \field$ generates a symmetry
of equation (\ref{eq1}) if $[F,G]=0$.

For any $a\in \field$ the time derivative $D_t(a)$ can be represented as
\begin{equation}\label{Dta}
 D_t (a)=a_* (F)\, .
\end{equation}
The variational derivative is defined as
\[
\frac{\delta a}{\delta u}=\sum_k (-1)^k D^k \left(\frac{\partial a}
{\partial u_k}\right)\, .
\]

Formal pseudo-differential series, which for simplicity we shall
call formal series, are defined as
\begin{equation}\label{A}
 A=a_{m}D^m+a_{m-1}D^{m-1}+\cdots + a_0+a_{-1}D^{-1}+a_{-2}D^{-2}+\cdots\,
\quad a_k\in \field\, .
\end{equation}
The product of two formal series is defined by
\begin{equation}\label{aDbD}
 aD^k\circ bD^m =a(bD^{m+k}+C_k^1 D(b)D^{k+m-1}+C_k^2 D^2 (b)D^{k+m-2}+
\cdots )\, ,
\end{equation}
where $k,m\in \mathbb Z$ and the binomial coefficients are
defined as
\[
C^j_n=\frac{n(n-1)(n-2)\cdots(n-j+1)}{j!}\,  .
\]
This product is associative.

\begin{Def} The formal series
\begin{equation}\label{Lm}
\Lambda=l_{m}D^m+l_{m-1}D^{m-1}+\cdots +
 l_0+l_{-1} D^{-1} +l_{-2} D^{-2}+ \cdots\,  , \quad l_k\in \field
\end{equation}
is called a formal recursion operator for equation (\ref{eq1}) if
\begin{equation}\label{Leq}
D_t(\Lambda)=F_*\circ \Lambda-\Lambda\circ F_*\,  .
\end{equation}
\end{Def}
In literature the formal recursion operator is also called the formal
symmetry of equation (\ref{eq1}).

The central result of the Symmetry Approach can be represented by the
following Theorem, which we attribute to A.Shabat (details of the proof
and applications one can find in \cite{ish,ss,msy,mss}):

\begin{The} If equation (\ref{eq1}) has an infinite hierarchy of higher
symmetries, then there exists a formal recursion operator.
\end{The}

The Theorem states that for integrable equations, i.e. equations
possessing an infinite hierarchy of higher symmetries, one can
solve equation (\ref{Leq}) and determine recursively the
coefficients $l_m,l_{m-1}, \ldots$ of  $\Lambda$ such that all
these coefficients will belong to the field $\field$. The
solvability conditions of equation (\ref{Leq}) can be formulated
in the elegant form of a canonical sequence of local conservation
laws of equation (\ref{eq1}), they provide powerful necessary
conditions of integrability. These conditions can be used for
testing for integrability  for a given equation or even for a
complete description of integrable equations of a particular
order.

\section{Differential polynomials and symbolic representation}

In what follows we shall consider equations (\ref{eq1}) which
right hand side is a differential polynomial or can be
represented in the form of a series

\begin{equation}\label{F}
F(u_{n},...,u_1, u_0)=F_1 [u]+F_2 [u]+F_3 [u]+\cdots \, ,
\end{equation}
where $F_k [u]$ is a homogeneous differential polynomial, i.e a
polynomial of variables $u_{n},...,u_1, u_0$ with complex constant
coefficients satisfying the condition $F_k[\lambda u]=\lambda^k
F_k[u], \lambda \in {\mathbb C}$, linear part $F_1[u]=L(u_0)$ and
$L$ is a linear operator ($\ord(L)\ge2$)

\begin{equation}\label{LL}
L=\sum_{k=0}^n r_k D^k\, , \quad r_k\in {\mathbb C}.
\end{equation}
For such equations we develop here a perturbative method to construct
formal recursion operator and testing for integrability. For simplicity
we shall consider the case when function $F$ is a differential
polynomial, i.e. the series (\ref{F}) contains a finite number of
terms. The generalisation to the case of infinite series will be
obvious.

Differential polynomials over ${\mathbb C}$ form a differential
ring $\ring$ which have a natural gradation
\begin{equation}\label{ring}
 \ring=\bigoplus_{n\ge 1}{\cal R}_n (u,D) \, ,
\end{equation}
where ${\cal R}_n (u,D)$ is a set of homogeneous differential
polynomials of degree $n$. The condition $n\ge 1$ in (\ref{ring}) means
that $1\not\in \ring$. In order to develop a perturbation theory and
for further generalisation of the approach to non-local cases it is
convenient to introduce a symbolic representation of this ring.

Symbolic representation (or symbolic method) was used in
mathematics since middle of nineteenth century. It was
successfully applied to the theory of integrable equations by
I.M. Gelfand and L.A. Dickey \cite{gd} in 1975 and also by V.E.
Zakharov and E.I. Schulman \cite{zsh}. Recently the power of this
method have been demonstrated again in the series of works of
J.Sanders and Jing Ping Wang (see for example \cite{w},
\cite{sw}) where they have given ultimate description of
integrable hierarchies of polynomial homogeneous evolution
equations.

Actually the symbolic representation is a simplified form of
notations and rules for formal Fourier images of dynamical
variables $u_n$, differential polynomials and formal series
(\ref{A}) with coefficients from the ring $\ring\oplus \mathbb C$.

Let $\hat{u}(\kappa ,t)$ denotes a Fourier image of $u(x,t)$
\[
u(x,t)=\int_{-\infty}^{\infty}\hat{u}(\kappa ,t)\exp (i\kappa
x)\,d\kappa \, ,
\]
then we have the following correspondences:
$ u_0\rightarrow \hat{u},\ u_1\rightarrow i\kappa\hat{u},\
\ldots \, \ u_m\rightarrow (i\kappa )^m \hat{u},\ ,\ldots$.
The Fourier image of a monomial $u_n u_m$ can obviously be
represented as
\begin{equation}\label{unum1}
u_n u_m=\int\!\!\!\int\!\!\!\int\delta(\kappa_1+\kappa_2-\kappa)
(i\kappa_1)^n (i\kappa_2)^m\hat{u}(\kappa_1 ,t)\hat{u}(\kappa_2 ,t)
\exp (i\kappa x)\,d\kappa_1 \,d\kappa_2\,d\kappa\end{equation}
and can be rewritten in a symmetrised form
\begin{equation}\label{unum2}
u_n u_m=\int\!\!\!\int\!\!\!\int\delta(\kappa_1+\kappa_2-\kappa)
\frac{[(i\kappa_1)^n (i\kappa_2)^m+(i\kappa_2)^n (i\kappa_1)^m]}{2}
\hat{u}(\kappa_1 ,t)\hat{u}(\kappa_2 ,t)
\exp (i\kappa x)\,d\kappa_1 \,d\kappa_2\,d\kappa \, ,
\end{equation}
therefore
\[
 u_n u_m\rightarrow\delta(\kappa_1+\kappa_2-\kappa)
\frac{[(i\kappa_1)^n (i\kappa_2)^m+(i\kappa_2)^n
(i\kappa_1)^m]}{2}\hat{u} (\kappa_1 ,t)\hat{u}(\kappa_2 ,t)\, .\]
We shall simplify notations further omitting the delta function,
replacing $i\kappa_n$ by $\xi_n$ and $\hat{u} (\kappa_1
,t)\hat{u}(\kappa_2 ,t)$ by $u^2$. Thus we shall represent the
monomial

$ u_n u_m$ by a symbol $u^2 a( \xi_1 , \xi_2 )$ where
\[ a(\xi_1,\xi_2)=\frac{[\xi_1^n \xi_2^m+\xi_2^n \xi_1^m]}{2} \]
is a symmetrical polynomial of its arguments. Following this rule we
shall represent any differential monomial $u_0^{n_0}u_1^{n_1}\cdots
u_q^{n_q}$ by the symbol
\[ u_0^{n_0}u_1^{n_1}\cdots u_q^{n_q}\rightarrow
u^{m}\langle \xi_1^0\cdots \xi_{n_0}^0\xi_{n_0+1}^1\cdots
\xi_{n_0+n_1}^1\xi_{n_0+n_1+1}^2\cdots \xi_{n_0+n_1+n_2}^2\cdots
\xi_{m}^q\rangle
\]
where  $m=n_0+n_1+\cdots +n_q$ and the brackets $\langle\rangle$
mean the symmetrisation over the group of permutation of $m$
elements (i.e. permutation of all arguments $\xi_j$)
\[
\langle f(\xi_1, \xi_2, \ldots ,
\xi_m)\rangle=\frac{1}{m!}\sum_{\sigma\in
\Sigma_m}f(\sigma(\xi_1),\sigma(\xi_2),\ldots ,\sigma(\xi_m))\, .
\]

For example
\[
u_n\rightarrow u \xi_1^n\,  ,\quad u_3^2\rightarrow u^2
\xi_1^3\xi_2^3\, ,\quad u^3 u_2\rightarrow
u^4\frac{\xi_1^2+\xi_2^2+\xi_3^2+\xi_4^2}{4}\, .
\]

We want to emphasise that the symmetrisation over the permutation
group is important, it is the symmetrisation makes the symbol
defined uniquely.  Equality of symbols implies the equality of the
corresponding differential polynomials.

The symbolic representation $\sring$ of the differential ring
$\ring$ can be defined as follows.  The sum of differential
monomials is represented by the sum of the corresponding symbols.
To the multiplication of monomials $f$ and $g$ with symbols $f\to
u^p a(\xi_1,\ldots ,\xi_p)$ and $g\to u^q b(\xi_1,\ldots ,\xi_q)$
corresponds the symbol
\[
fg\to u^{p+q}\langle a(\xi_1,\ldots ,\xi_p)b(\xi_{p+1},\ldots
,\xi_{p+q})\rangle \, .
\]
Here the symmetrisation is taken over the group of permutation of
all $p+q$ arguments $\xi_1,\ldots \xi_{p+q}$.  The derivative
$D(f)$ of a monomial  $f$ with the symbol $u^p a(\xi_1,\ldots
,\xi_s)$ is represented by
\[ D(f)\to u^s (\xi_1+\xi_2+\cdots +\xi_p)a(\xi_1,\ldots ,\xi_s)\, .\]

The following rules are motivated by the theory of linear
pseudo-differential operators in Fourier representation and are
nothing but abbreviated notations.  To the operator $D$ (\ref{D})
we shall assign a special symbol $\eta$ and the following rules of
action on symbols:

\[ \eta( u^n a(\xi_1,\ldots , \xi_n))=u^n a(\xi_1,\ldots , \xi_n)\sum_{j=1}^{n}\xi_j \]
and the composition rule
\[
\eta\circ u^n a(\xi_1,\ldots , \xi_n)=u^n a(\xi_1,\ldots ,
\xi_n)(\sum_{j=1}^{n}\xi_j+\eta )\, .
\]
The latter corresponds to the Leibnitz rule $D\circ f=D(f)+f D$.
Now it can be shown that  the composition rule (\ref{aDbD}) can be
represented as following. Let we have two operators $fD^q$ and
$gD^s$ such that  $f$ and $g$ have symbols $u^i a(\xi_1,\ldots
,\xi_i)$ and $u^j b(\xi_1,\ldots ,\xi_j)$ respectively. Then
$fD^q\to u^i a(\xi_1,\ldots ,\xi_i)\eta^q,gD^s\to u^j
b(\xi_1,\ldots ,\xi_j)\eta^s $ and
\begin{equation}\label{fDgD}
fD^q \circ gD^s\to u^{i+j}\langle a(\xi_1,\ldots ,\xi_i)
(\eta+\sum_{m=i+1}^{i+j}\xi_m)^q b(\xi_{i+1},\ldots
,\xi_{i+j})\eta^s \rangle\, .
\end{equation}
Here the symmetrisation is taken over the group of permutation of
all $i+j$ arguments $\xi_1,\ldots \xi_{i+j}$, the symbol $\eta$ is
not included in this set.  In particularly it follows from
(\ref{fDgD}) that $D^q\circ D^s\to \eta^{q+s}$.  The composition
rule (\ref{fDgD}) is valid for positive and negative exponents
$q,s$. In the case of positive exponents it is a polynomial in
$\eta$ and the result is a Fourier image of a differential
operator. In the case of negative exponents one can expand the
result on $\eta$ at $\eta\to \infty$ in order to identify it with
(\ref{aDbD}). In the symbolic representation instead of formal
series (\ref{A}) it is natural to consider formal series of the
form
\begin{equation}\label{B}
B=b(\eta)+u b_1(\xi_1,\eta)+u^2 b_2(\xi_1,\xi_2,\eta)+u^3
b_3(\xi_1,\xi_2,\xi_3,\eta)+\cdots\, ,\quad b(\eta)\not = 0
\end{equation}

Let $fD^q\to u^i a(\xi_1,\ldots ,\xi_i)\eta^q$ then the symbolic
representation for the formally conjugated operator is

\[(-1)^qD^q\circ f\to u^i a(\xi_1,\ldots ,\xi_i)(-\eta-\sum_{n=1}^i \xi_n)^q\, .\]

The symbolic representation of the Fr\'echet derivative of the
element $f\to u^n a(\xi_1,\ldots ,\xi_n)$ is
\[ f_*\to nu^{n-1}a(\xi_1,\ldots ,\xi_{n-1},\eta )\, .\]
For example, let $F=u_3+6uu_1$, then $F\to u \xi_1^3+3 u^2
(\xi_1+\xi_2)$ and
\[ F_*\to \eta^3+6u (\xi_1+\eta )\, .\]
It is interesting to notice that the symbol of the Fr\'echet
derivative is always symmetric with respect to all permutations of
arguments, including the argument $\eta$. Moreover, the following
obvious, but useful Proposition hold:

\begin{Pro} A differential operator is a Fr\'echet derivative of an element
of $\ring$ if and only if its symbol is invariant with respect to
all permutations of its argument, including the argument $\eta$.
\end{Pro}

The variational derivative $\delta f/\delta u$ of $f\to u^n
a(\xi_1,\ldots ,\xi_m)$ can be represented as
\[
\frac{\delta f}{\delta u}\to mu^{m-1}a(\xi_1,\ldots ,\xi_{m-1},
-\sum_{i=1}^{m-1} \xi_i)\, .
\]

The symbolic representation has been extended and proved to be
very useful in the case of noncommutative differential rings
\cite{ow}. It can be easily generalised to the case of many
dependent variables \cite{swsys}, suitable for study of system of
equations. Here we are going to extend it further to the case of
non-local and multidimensional equations.

\section{Symmetry Approach in symbolic representation}

Let the right hand side of equations (\ref{eq1}) be a differential
polynomial or can be represented in the form of a series
(\ref{F}). In the symbolic representation it can be written as
\begin{equation}
\label{eqsym} u_t=u\omega(\xi_1)+\frac{u^2}{2}a_1
(\xi_1,\xi_2)+\frac{u^3}{3}a_2 (\xi_1,\xi_2,\xi_3)+\frac{u^4}
{4}a_3 (\xi_1,\xi_2,\xi_3,\xi_4)+\cdots=F\, ,
\end{equation}
where $\omega (\xi_1), a_n (\xi_1,...,\xi_{n+1})$ are symmetrical
polynomials and $\deg \omega (\xi_1)\ge 2$.  According the
previous section the Fr\'echet derivative of the right hand side
is of the form
\begin{equation}
\label{F*} F_*=\omega(\eta )+ua_1 (\xi_1,\eta )+u^2a_2
(\xi_1,\xi_2,\eta )+u^3a_3(\xi_1,\xi_2,\xi_3,\eta )+\cdots\ .
\end{equation}

Symmetries of equation (\ref{eqsym}), if they exsist, can be found recursively:

\begin{Pro}\label{prosymmetry}
Suppose the equation (\ref{eqsym}) has a symmetry
\begin{equation}
\label{Geq} u_{\tau}=u\Omega(\xi_1)+\sum_{j\ge
1}\frac{u^{j+1}}{j+1}A_j(\xi_1,\ldots,\xi_{j+1})=G
\end{equation}
Then functions $A_j(\xi_1,\ldots,\xi_{j+1})$ of the symmetry are
related to functions $a_i(\xi_1,\ldots,\xi_{i+1})$ of the equation
by the following formulae
\begin{eqnarray}\label{A1}
&&A_1(\xi_1,\xi_2)=\frac{N^\omega (\xi_1,\xi_2)} {N^\Omega
(\xi_1,\xi_2)}a_1(\xi_1,\xi_2)\\ \label{Am}
&&A_m(\xi_1,\ldots,\xi_{m+1})=\frac{N^\omega
(\xi_1,...,\xi_{m+1})}{N^\Omega(\xi_1,...,\xi_{m+1})}a_m(\xi_1,\ldots,\xi_{m+1})+\\
\nonumber &&+N^\omega
(\xi_1,...,\xi_{m+1})\langle\sum_{j=1}^{m-1}\frac{m+1}{m-j+1}
A_j(\xi_1,\ldots,\xi_j,
\xi_{j+1}+\cdots+\xi_{m+1})a_{m-j}(\xi_{j+1},\ldots,\xi_{m+1})
\\ \nonumber
&&-\sum_{j=1}^{m-1}\frac{m+1}{j+1} a_{m-j} (\xi_1,\ldots,
\xi_{m-j}\xi_{m-j+1}+
\cdots+\xi_{m+1})A_j(\xi_{m-j+1},\ldots,\xi_{m+1}) \rangle
\end{eqnarray}
where
\begin{equation}\label{Nomega}
N^\omega
(\xi_1,...,\xi_m)=\left(\omega(\sum_{n=1}^{m}\xi_n)-\sum_{n=1}^{m}\omega(\xi_n)\right)^{-1},
N^\Omega
(\xi_1,...,\xi_m)=\left(\Omega(\sum_{n=1}^{m}\xi_n)-\sum_{n=1}^{m}\Omega(\xi_n)\right)^{-1}\,
.
\end{equation}
\end{Pro}

{\bf Proof} To find a symmetry we have to solve equation $F_*(G)=G_*(F)$ with respect to $G$.
 For $F_*(G)$ we have
\[
F_*(G)=u\omega(\xi_1)\Omega(\xi_1)+\sum_{j\ge
1}\frac{u^{j+1}}{j+1}\omega(\xi_1+\cdots+\xi_{k+1})A_j(\xi_1,\ldots,\xi_{j+1})+
\]
\[
+\sum_{i\ge 1}\frac{u^{i+1}}{i+1} a_i(\xi_1,\ldots,\xi_{i+1})
[\Omega(\xi_1)+ \cdots+ \Omega(\xi_{i+1})] +
\]
\[
+\sum_{i\ge 1}\sum_{j\ge 1}\frac{u^{i+j+1}}{j+1}\langle a_i(\xi_1,
\ldots,\xi_i, \xi_{i+1}+\cdots+ \xi_{i+j+1})A_j(\xi_{i+1}, \ldots,
\xi_{i+j+1})\rangle
\]
Similarly we obtain
\[
G_*(F)=u\omega(\xi_1)\Omega(\xi_1)+\sum_{i\ge 1}\frac{u^{i+1}}
{i+1} \Omega(\xi_1 +\cdots+\xi_{i+1})a_i(\xi_1,\ldots,\xi_{i+1})+
\]
\[
\sum_{j\ge 1}\frac{u^{j+1}}{j+1}A_j(\xi_1,
\ldots,\xi_{j+1})[\omega(\xi_1) +\cdots+\omega(\xi_{j+1})]+
\]
\[
+\sum_{i\ge 1}\sum_{j\ge 1}\frac{u^{i+j+1}}{i+1}\langle
A_j(\xi_1,\ldots,\xi_j,\xi_{j+1}+\cdots+\xi_{i+j+1})a_i(\xi_{j+1},\ldots,\xi_{i+j+1})\rangle
\]
Substituting that relations into the equation $F_*(G)=G_*(F)$ and
collecting the coefficients at each power of $u^j$ we find simple
equations to determine $A_m(\xi_1,...,\xi_{m+1})$. Solutions of
these equations are given by (\ref{A1}), (\ref{Am}).
$\quad_{\triangle}$

For any function $F$ of the form (\ref{eqsym}) we can solve the linear
operator equation (\ref{Leq}) find a formal recursion operator
$\Lambda$.

\begin{Pro}\label{prolambda}
Operator $\Lambda$ is a solution of equation (\ref{Leq}) if its symbol is
of the form
\begin{equation}
\label{Lsym} \Lambda=\phi(\eta )+u\phi_1 (\xi_1,\eta )+u^2\phi_2
(\xi_1,\xi_2,\eta )+u^3\phi_3 (\xi_1,\xi_2,\xi_3,\eta )+\cdots \,
,
\end{equation}
where $\phi(\eta )$ is an arbitrary function and
$\phi_m(\xi_1,...,\xi_m,\eta )$ are determined recursively
\begin{eqnarray}\label{phi1}
&&\phi_1 (\xi_1,\eta )=N^\omega(\xi_1,\eta ) a_1(\xi_1,\eta )
(\phi(\eta+\xi_1)-\phi(\eta ))\\  \label{phim} &&\phi_m
(\xi_1,...,\xi_m,\eta )=N^\omega (\xi_1,...,\xi_m,\eta ) \{
(\phi(\eta+\xi_1+\cdots +\xi_m)-\phi(\eta )) a_m(\xi_1,...,\xi_m,\eta )+\\
\nonumber && \sum_{n=1}^{m-1}\langle \frac{n}{m-n+1}
\phi_n(\xi_1,...,\xi_{n-1},\xi_n+\cdots +\xi_m,\eta ) a_{m-n}(\xi_n,...,\xi_m)+\\
\nonumber && \phi_n(\xi_1,...,\xi_{n},\eta+\xi_{n+1}+\cdots
+\xi_m) a_{m-n}(\xi_{n+1},...,\xi_m,\eta )-\\ \nonumber &&
a_{m-n}(\xi_{n+1},...,\xi_m,\eta+\xi_{1}+\cdots +\xi_n)
\phi_n(\xi_{1},...,\xi_n,\eta ) \rangle \}
\end{eqnarray}
\end{Pro}

{\bf Proof}
Using (\ref{Dta}) we find
\begin{eqnarray*}
D_t(\Lambda)&=& \sum_{n\ge 1}u^n\phi_n (\xi_1,...,\xi_n,\eta
)\sum_{i=1}^n \omega(\xi_i)+\\&& \sum_{n\ge 1,m\ge
1}\frac{n}{m+1}u^{n+m}\langle \phi_n(\xi_1,...,\xi_{n-1},
\xi_n+\xi_{n+1}+\cdots +\xi_{n+m},\eta )
a_m(\xi_n,...,\xi_{n+m})\rangle\, .
\end{eqnarray*}
It follows from the composition rule (\ref{fDgD}) that
\begin{eqnarray*}
[F_*, \Lambda ]&=&\sum_{m\ge 1}u^m\left( (\omega
(\eta+\xi_1+\cdots +\xi_m)- \omega (\eta ) )\phi_m
(\xi_1,...,\xi_m,\eta ) \right. -
\\
& & \left. (\phi(\eta+\xi_1+\cdots +\xi_m)-\phi(\eta ))a_m (\xi_1,...,\xi_m,\eta )\right) + \\
& & \sum_{n\ge 1,m\ge 1}u^{n+m}\langle a_n
(\xi_1,...,\xi_n,\eta+\xi_{n+1}+
\cdots +\xi_{n+m}) \phi_m (\xi_{n+1},...,\xi_{n+m},\eta ) - \\
& & \phi_m (\xi_1,...,\xi_m,\eta+\xi_{m+1}+\cdots \xi_{m+n}) a_n
(\xi_{m+1},...,\xi_{m+n},\eta )\rangle\, .
\end{eqnarray*}
Collecting the coefficients at each power $u^m$ we find that the
structure of the relations is triangular, and all coefficients  $\phi
_m$ can be found recursively. $\quad_{\triangle}$

We immediately see the advantage of the perturbative approach.
Now we are able to obtain explicit recursion relations for
determining the coefficients of a symmetry and a formal recursion
operator while in the standard Symmetry Approach the
corresponding problem was quite difficult.

Existence of a symmetry means that all coefficients
$A_m(\xi_1,\ldots,\xi_{m+1})$ are polynomials (not rational
functions). In other words the symbols $u^{m+1}
A_m(\xi_1,\ldots,\xi_{m+1})\in\sring$ and correspond to
differential polynomials in the standard representation. This
requirement can be used for testing for integrability and even
for complete classification of integrable equations (see
\cite{sw}, \cite{w}, \cite{ow}).

In the standard Symmetry Approach the integrability, i.e. the
existence of infinite hierarchies of local symmetries or
conservation laws implies (Theorem 1) that all coefficients $l_n$
are local and belong to the corresponding differential field or
ring. In the symbolic representation it suggests the following
definition.

\begin{Def}
 We say that the function $b_m
(\xi_1,...,\xi_m,\eta), \ m\ge 1$ is $k$-local if the first $k$
coefficients $\beta_{mn} (\xi_1,...,\xi_m),\ n=n_s,...,n_s+k$ of
its expansion at $\eta\to\infty$
\[
b_m (\xi_1,...,\xi_m,\eta)=\sum_{n=s_n}^\infty \beta_{mn}
(\xi_1,...,\xi_m)\eta^{-n}
\]
are symmetric polynomials. We say
that the coefficient $b_m (\xi_1,...,\xi_m,\eta)$ of a formal
series (\ref{B}) is local if it is $k$-local for any $k$.
\end{Def}

\begin{The}
Suppose equation (\ref{eqsym}) has an infinite hierarchy of
symmetries
\begin{equation}
\label{symi} u_{t_i}=u\Omega_i(\xi_1)+\sum_{j\ge
1}\frac{u^{j+1}}{j+1}A_{ij}(\xi_1,\ldots,\xi_{j+1})=G_i\, ,\quad
i=1,2,\ldots
\end{equation}
where $\Omega_i(\xi _1)$ are polynomials of degree
$m_i=\deg(\Omega_{i}(\xi_1))$ and $m_1 <m_2<\cdots < m_{i}<\cdots
$. Then the coefficients $\phi_m(\xi_1,...,\xi_m,\eta)$ of the
formal recursion operator
\begin{equation}
\label{lp} \Lambda=\eta+u\phi_1(\xi_1,\eta
)+u^2\phi_2(\xi_1,\xi_2,\eta )+\cdots
\end{equation}
are local.
\end{The}

{\bf Proof.} We use the following obvious Lemma.

\begin{Lem} For any formal series of the form (\ref{lp}) let $P(\Lambda)$
be a polynomial of $\Lambda$ (of a degree $q\ge 1$) with constant
coefficients, then the first $N$ terms of the formal series
$P(\Lambda)$ are $k$-local if and only if the first $N$ terms of
$\Lambda$ are $k$-local.
\end{Lem}

In order to prove the theorem we will show by induction that for
any $N$ and $k$ there exist $q$ such that $N$ first coefficients
$\phi_1(\xi_1,\eta),...,\phi_N(\xi_1,...,\xi_N,\eta)$ of the
formal recursion operator
\[ \Lambda _q=\Omega_q(\Lambda)=\Omega_q(\eta)+u\tilde{\phi}_1(\xi_1,\eta
)+u^2\tilde{\phi}_2(\xi_1,\xi_2,\eta )+\cdots
\]
are $k$-local.

 It follows from (\ref{A1}), (\ref{Nomega}) that
\[
A_{q1}(\xi_1,\xi_2)=\frac{\Omega_q(\xi_1+\xi_2)-\Omega(\xi_2)}{\omega(\xi_1+\xi_2)-
\omega(\xi_1)- \omega(\xi_2)}a_1(\xi_1,\xi_2)-
\frac{\Omega(\xi_1)} {\omega(\xi_1+\xi_2)- \omega(\xi_1)-
\omega(\xi_2)}a_1(\xi_1,\xi_2)
\]
Recalling (\ref{phi1}) we obtain
\begin{equation}
\label{A1ex} \tilde\phi_1(\xi_1,\eta )=A_{q1}(\xi_1,\eta
)+R_1(\xi_1,\eta )
\end{equation}
where
\[
R_1(\xi_1,\eta )=\frac{\Omega(\xi_1)}{\omega(\xi_1+\eta
)-\omega(\xi_1)- \omega(\eta )}a_1(\xi_1,\eta )
\]
Expanding the last two formulae on $1/\eta$ and taking into
account that $A_{q1}(\xi_1,\eta)$ is a polynomial and therefore a
local function and
\[
\tilde\phi_1(\xi_1,\eta )=\sum_{j\le
s_1}\tilde\phi_{1j}(\xi_1)\eta^j,\quad R_1(\xi_1,\eta )=\sum_{j\le
l1}r_{1j}(\xi_1)\eta^j
\]
where
\[
s_1=m_q-n+n_1,\quad l_1=n_1-n+1\, .
\]
We see that at least first $s_1-l_1-1=m_q-2$ coefficients
$\tilde\phi_{1j}(\xi_1)$ are polynomials and therefore
$\tilde\phi_1(\xi_1,\eta)$ is $k$-local and $k=m_q-2$. Thus for
any $k$ there exists $m_q$ such that the first coefficient
$\phi_1(\xi_1,\eta)$ of $\Lambda$ is $k$-local.

Let us assume that there exists $m_q$ such that first $j-1$
coefficients $\phi_{n}(\xi_1,...,\xi_{n},\eta),\ n=1,...,j-1$ are
$k$-local, then we will show that the coefficient
$\phi_{j}(\xi_1,...,\xi_{j},\eta)$ is $k$-local for sufficiently
large $m_q$.

It follows from (\ref{Am}), (\ref{phim}) that the coefficient
$\tilde\phi_j(\xi_1,\ldots,\xi_j,\eta )$ can be represented as
\begin{equation}
\tilde\phi_j(\xi_1,\ldots,\xi_j,\eta )= \label{Ajex}
A_j(\xi_1,\ldots,\xi_j,\eta )+ R_j(\xi_1, \ldots, \xi_j, \eta )
\end{equation}
where
\[
R_j(\xi_1,\ldots,\xi_j,\eta )= -N^\omega
(\xi_1,...,\xi_j,\eta)\bigg[\sum_{i=1}^{j-1}\bigg( \langle
R_i(\xi_1,\ldots,\xi_i,\xi_{i+1}+\cdots+\xi_j+\eta )\cdot
a_{j-i}(\xi_{i+1},\ldots,\xi_j,\eta )\rangle-\]
\[\langle a_{j-i}
(\xi_1, \ldots, \xi_{j-i}, \xi_{j-i+1} + \cdots+\xi_j+\eta )
R_i(\xi_{i+1}, \ldots,\xi_{j},\eta ) \rangle+
\]
\[
\frac{i}{j-i+1}\langle
R_i(\xi_1,\ldots,\xi_{i-1},\xi_i+\cdots+\xi_j,\eta
)a_{j-i}(\xi_i,\ldots,\xi_j,)\rangle-
\]
\[
\frac{j-i}{i+1}\langle
a_{j-i}(\xi_1,\ldots,\xi_{j-i-1},\xi_{j-i}+\cdots+\xi_j,\eta
)A_i(\xi_{j-i}, \ldots,\xi_j) \rangle
\bigg)-(\Omega_q(\xi_1)+\cdots+\Omega(\xi_j))a_j(\xi_1,
\ldots,\xi_j,\eta ) \bigg]
\]

Taking the expansion of (\ref{Ajex}) in $1/\eta$ find that the
principal power of $\eta$ for  $\phi_j$ can be represented as
$m_q+S(n,n_1,...,n_j)$ and the principal power for $R_j$ as
$Q(n,n_1,...,n_j)$, therefore for sufficiently large $m_q$ we
will have $m_q+S(n,n_1,...,n_j)-Q(n,n_1,...,n_j)>k$, i.e. the
coefficient $\phi_j$ is $k$-local.

For any $N$ and $k$ there exists such $m_q$ that first $N$
coefficients of $\Lambda$ are $k$-local. $\quad_{\triangle}$

The symmetry approach in symbolic representation suggests the following
test for integrability of equations of the form (\ref{eqsym}):
\begin{itemize}
\item Find a first few coefficients $\phi_n(\xi_1,...,\xi_n,\eta )$.
\item Expand these coefficients in series of $1/\eta$
\begin{equation}\label{Phi}
\phi_n(\xi_1,...,\xi_n,\eta
)=\sum_{s=s_n}\Phi_{ns}(\xi_1,...,\xi_n)\eta^{-s}
\end{equation}
and find the corresponding functions $\Phi_{ns}(\xi_1,...,\xi_n)$.
\item Check that functions $\Phi_{ns}(\xi_1,...,\xi_n)$ are polynomials (not rational functions).
\end{itemize}
First three nontrivial coefficients $\phi_n$ were sufficient to analyse
in all known to us cases.

As an example of application we consider equations of the form
\begin{equation}
\label{eqex} u_t=u \omega(\xi_1)+\sum_{i\ge
1}\frac{u^{i+1}}{i+1}a_i(\xi_1,\ldots,\xi_{i+1}),
\end{equation}
where $\omega(\xi_1)$ is a polynomial on $\xi_1$ of the degree
$deg(\omega(\xi_1))=n\ge 2$ and $a_i(\xi_1,\ldots,\xi_{i+1})$ are
symmetric polynomials on its arguments of degree
$deg(a_i(\xi_1,\ldots,\xi_{i+1}))\le n-2$. The following
propositions are valid.

\begin{Pro} \label{pro-omega}
If equation (\ref{eqex}) is integrable, then $\omega (0)=0$, i.e.
the polynomial $\omega(\xi_1)$ can be factorised
$\omega(\xi_1)=\xi_1 f(\xi_1)$, where $f(\xi_1)$ is a polynomial.
\end{Pro}

\begin{Pro}\label{pro-non}
Suppose $n$ is even and $a_1(\xi_1,\xi_2)\equiv 0$ or $n$ is odd
and $a_1(\xi_1,\xi_2)\equiv 0,a_2(\xi_1,\xi_2,\xi_3)\equiv 0$.
Then  equation (\ref{eqex}) is not integrable.
\end{Pro}

{\bf Proof of proposition 4} Let us consider first the case of
even $n$.  Let we have the equation of the form (\ref{eqex}),
where $a_1(\xi_1,\xi_2)=\ldots= a_{s-1}(\xi_1,\ldots,\xi_s)=0,
a_s(\xi_1,\ldots,\xi_{s+1})\ne 0$. For corresponding $\Lambda$ -
operator using (\ref{phi1}), (\ref{phim}) we have
\[
\Lambda=\eta+\sum_{j\ge s}u^j\phi_j(\xi_1,\ldots,\xi_j,\eta ).
\]
Let us consider coefficient
\[
\phi_s(\xi_1,\ldots,\xi_s,\eta
)=\frac{(\xi_1+\cdots+\xi_s)a_s(\xi_1,\ldots,\xi_s,\eta )}
{\omega(\xi_1+\cdots+\xi_s+\eta )-\omega(\eta
)-\omega(\xi_1)-\cdots-\omega(\xi_s)}
\]
For $s\ge 2$ the polynomial $\omega(\xi_1+\cdots+\xi_s+\eta
)-\omega(\eta )- \omega(\xi_1)-\cdots-\omega(\xi_s)$ is of degree
$n$ and cannot be factorised \cite{w}. Therefore it cannot divide
the numerator which is a factorisable polynomial of degree $n-1$.

This coefficient is nonlocal. Indeed, the expansion of denominator
$\omega(\xi_1+\cdots+\xi_s+\eta )-\omega(\eta
)-\omega(\xi_1)-\cdots-\omega(\xi_s)$ at $\eta\to\infty$ is of the
form
\[
[\omega(\xi_1+\cdots+\xi_s+\eta )-\omega(\eta
)-\omega(\xi_1)-\cdots-\omega(\xi_s)]^{-1}=
\]
\[
=\frac{1}{\omega'(\eta )(\xi_1+\cdots+\xi_s)}\sum_{j\ge
0}\bigg[\frac{\omega(\xi_1)+ \cdots+\omega(\xi_s)} {\omega'(\eta
)(\xi_1 +\cdots+\xi_s)}- \sum_{i=2}^n\frac{\omega^{(i)}(\eta )}
{\omega'(\eta )} \frac{(\xi_1+ \cdots+\xi_s)^{i-1}}{i!} \bigg]^j
\]
and it contains powers of a singular term
$\frac{\omega(\xi_1)+\cdots+ \omega(\xi_s)} {(\xi_1+
\cdots+\xi_s)}$ (for $s\ge 1$ and even $n$ the sum
$\omega(\xi_1)+\cdots+\omega(\xi_{s})\not \equiv 0$ on the
hyperplane $\xi_1+\cdots +\xi_s=0$ .

In the case of odd $n$ the proof is analogous with the only
difference that the function $\omega(\xi_1+\cdots+\xi_s+\eta
)-\omega(\eta )-\omega(\xi_1)-\cdots-\omega(\xi_s)$ is not
factorisable if $s\ge 3$ and $\omega(\xi_1)+\omega(-\xi_1)$ may
be equal zero for some dispersion laws $\omega(k)$. Proposition
is proved. $\quad_{\triangle}$

{\bf Proof of proposition 3} We consider polynomial dispersion law
$\omega(k)$ and we need to proof that $\omega(0)=0$. Let the
degree of the $\omega(k)$ $n$ be odd. As it follows from
Proposition 4 if equation (\ref{eqex}) is integrable then either
$a_1(\xi_1,\xi_2)\ne 0$ or $a_1(\xi_1,\xi_2)=0$, but
$a_2(\xi_1,\xi_2,\xi_3)\ne 0$. Let us consider coefficient
$\phi_1(\xi_1,\eta )$ of the corresponding $\Lambda$-operator in
the case $a_1(\xi_1,\xi_2)\ne 0$:
\[
\phi_1(\xi_1,\eta )=\frac{\xi_1a_1(\xi_1,\eta )}{\omega(\xi_1+\eta
)-\omega(\xi_1)-\omega(\eta )}
\]
Its expansion on $\eta$ at the point $\eta\to\infty$ is of the
form
\[
\phi_1(\xi_1,\eta )=\frac{a_1(\xi_1,\eta )}{\omega'(\eta
)}\sum_{j\ge 0} \bigg[ \frac{\omega(\xi_1)} {\xi_1\omega'(\eta )}-
\sum_{i=2}^n \omega^{(i)} (\eta ) \frac{\xi_1^i} {i!}\bigg]^j
\]
and contains singularity $\omega(\xi_1)/\xi_1$ in any power. If
$\omega(0)\ne 0$ then the polynomial $ \omega(\xi_1+\eta
)-\omega(\xi_1)-\omega(\eta )$ is not factorisable. It cannot
divide the numerator $\xi_1a_1(\xi_1,\eta )$ since $\deg
a_1(\xi_1,\eta )<\deg \omega(\eta)$.

The cases of even $n$ and
$a_1(\xi_1,\xi_2)=0,a_2(\xi_1,\xi_2,\xi_3)\ne 0$, odd $n$ can be
proved in a similar way.  $\quad_{\triangle}$

The statement of proposition 4 in the homogeneous case was proved
in works of J. Sanders and J.P. Wang \cite{sw}.

\section{Non-local extensions}

The main goal of the paper is to extend the symmetry approach to the
case of nonlocal and non-evolutionary equations. Multi-dimensional
integrable equations and their hierarchies, have also intrinsic
non-locality and certain modifications of the standard Symmetry
Approach  are required \cite{MY}. The symbolic representation seems to
be suitable to tackle the problem of non-locality.

Here we consider two types of non-local equations. The first one
is the Benjamin-Ono equation
\begin{equation}\label{BO}
u_t=H(u_2) + 2 uu_1
\end{equation}
where $H(f)$ denotes the Hilbert transform
\begin{equation}\label{H}
H(f)=\frac{1}{\pi}\int_{-\infty}^\infty \frac{f(y)}{y-x}\, dy\, .
\end{equation}
The second example is
\begin{equation}\label{CHD}
m_t=c m u_1 + u m_1\, ,\qquad m=u-u_2\, ,\quad c\ne 0\, ,
\end{equation}
where $c$ is a constant (this equation is known to be integrable for
$c=2$ \cite{CH} and $c=3$ \cite{D}, see also \cite{DHH}). Our approach
can be extended to the multidimensional case, where non-locality is an
intrinsic property (we are going to present the multi-dimensional case
in a separate paper).

Equation (\ref{BO}) is nonlocal (not differential), system (\ref{CHD})
is non-evolutionary. Their higher symmetries, when they exist are even
more nonlocal. To tackle the problem of non--locality a proper
extension of the differential ring is required. In the 2+1 dimensional
case such extension was proposed in \cite{MY}. Here we are developing
this idea and reformulate it in the frame of perturbative symmetry
approach and illustrate by examples.

\subsection{Benjamin-Ono type equations}

The Benjamin-Ono equation (\ref{BO}) contains the Hilbert
transform. It is well known that its higher symmetries and
conservation laws contain nested Hilbert transforms and we have
to extend the differential ring $\ring$ in order to study
structures associated with the Benjamin-Ono type equations.

The construction of the extension is quite similar to \cite{MY}. Let us
consider the following sequence of the ring extensions:
\begin{equation}\label{R^n}
{\cal R}^0_H=\ring\, , \quad {\cal R}^1_H=\overline{{\cal
R}^0_H\bigcup H({\cal R}^0_H)} \, ,\quad {\cal
R}^{n+1}_H=\overline{{\cal R}^n_H\bigcup H({\cal R}^n_H)}\, ,
\end{equation}
where the set $H({\cal R}^n_H)$ is defined as $H({\cal R}^n_H)=\{
H(a); a\in{\cal R}^n_H\}$ and the horizontal line denotes the ring
closure. Each ${\cal R}^n_H$ is a ring and the upper index $n$
indicates the nesting depth of the operator $H$.

Elements of ${\cal R}^0_H$ are differential polynomials. Elements
of ${\cal R}^n_H\, \ n\ge 1$ we call {\sl quasi--local functions}.
The right hand side of equation (\ref{BO}), its symmetries and
densities of conservation laws are quasi--local functions.

In the symbolic representation operator $H$ is represented by
$i\sign (\eta )$.  Now the symbolic representation of the ring
extensions is obvious.  Suppose element $A\in {\cal R}^0_H$ and
the corresponding symbol is $u^n a(\xi_1,...,\xi_n)$, then $H(A)$
is represented by the symbol $u^n b(\xi_1,...,\xi_n)$, where
$b(\xi_1,...,\xi_n)=i \sign (\xi_1+\cdots
+\xi_n)a(\xi_1,...,\xi_n)$.

In symbolic representation all definitions, such as the Fr\'eche
derivative and formal recursion operator are exactly the same as
in the local case.  Now we have more freedom in the choice of
function $\phi(\eta )$ (Proposition 3).  We can choose $\phi(\eta
)=\eta$ or $\phi(\eta )=\eta\sign (\eta )$. The asymptotic
locality conditions, i.e. the conditions that the coefficients
$u^n \Phi_{nm}(\xi_1,...,\xi_n)$ of the formal recursion operator
$\Lambda$ (\ref{Lsym}) represent symbols of differential
polynomials (and therefore $\Phi(\xi_1,...,\xi_n)$ are
symmetrical polynomials) now should be replaced by the
quasi--locality conditions. Namely, that the coefficients  $u^n
\Phi_{nm}(\xi_1,...,\xi_n)$ represent symbols of quasi--local
functions means that the coefficients  $u^n
\Phi_{nm}(\xi_1,...,\xi_n)$ correspond to elements from the
extended ring.

For example let us consider Benjamin-Ono equation (\ref{BO}).  In
symbolic representation it reads as follows
\begin{equation}
\label{BOs} u_t=i u \sign (\xi_1)\xi_1^2+u^2(\xi_1+\xi_2)
\end{equation}
The first coefficient $\phi_1(\xi_1,\eta )$ of corresponding
formal recursion operator $\Lambda=\eta+u\phi_1(\xi_1,\eta
)+u^2\phi_2(\xi_1,\xi_2,\eta )+\cdots$ looks as follows
\[
\phi_1(\xi_1,\eta )=\sign(\eta
)+\frac{\xi_1(\sign(\xi_1)+\sign(\eta ))}{2\eta}+O(\frac{1}{\eta^7})
\]
and it is evident that it is quasi-local. One may easily check
quasi-locality of other coefficients $\phi_2(\xi_1,\xi_2,\eta
),\phi_3(\xi_1,\xi_2,\xi_3,\eta ),...$.

As another illustration let us consider equation of the form
\begin{equation}
\label{BOg}
u_t=\hat{H}(u_2)+c_1uu_1+c_2\hat{H}(uu_1)+c_3u\hat{H}(u_1)+c_4u_1\hat{H}(u)+
\end{equation}
$$ +c_5\hat{H}(u \hat{H}(u_1))+ c_6\hat{H}(u)\hat{H}(u_1) $$ where
$c_j$ are complex constants.  Where $\hat{H}= -i H$ so the symbol
for $\hat{H}$ now reads as $\sign $). We announce here the
following theorem (details and proof see in \cite{miknov2})

\begin{The}
Equation of the form (\ref{BOg}) is integrable if and only if it
is up to the point transformation $u\to a u+b \hat{H}(u),\quad
a^2-b^2\ne 0$ one of the list
\begin{equation}
\label{eqA}
u_t=\hat{H}(u_2)+D(\frac{1}{2}c_1u^2+c_2u\hat{H}(u)+\frac{1}{2}c_1\hat{H}(u)^2)
\end{equation}
\begin{equation}
\label{eqB}
u_t=\hat{H}(u_2)+D(\frac{1}{2}c_1u^2+\frac{1}{2}c_2\hat{H}(u^2)-c_2u\hat{H}(u))
\end{equation}
\begin{equation}
\label{eqC} u_t=\hat{H}(u_2)+uu_1\pm \hat{H}(uu_1)\mp u
\hat{H}(u_1)\mp 2u_1\hat{H}(u)+\hat{H}(u \hat{H}(u_1))
\end{equation}
\begin{equation}
\label{eqD} u_t=\hat{H}(u_2)+\hat{H}(uu_1)+u_1\hat{H}(u)\pm
\hat{H}(u \hat{H}(u_1))\pm \hat{H}(u)\hat{H}(u_1)
\end{equation}
\end{The}

As the last example in this subsection let us consider equation of the
form
\begin{equation}
\label{Bodl} u_t=u\omega(\xi_1)+\frac{u^2}{2}(\xi_1+\xi_2).
\end{equation}
\begin{Pro} Let the dispersion relation $\omega(k)$  for equation
(\ref{Bodl}) be of the form $\omega(k)=k^2 f(k)$ where function
$f(k)\to 1$ faster than any power of $k$ when $k\to \infty$ and
$f(k)\to c_{-1} k^{-1}+c_0+c_1 k+c_2 k^2+\cdots $ when $k\to 0$. Then
equation (\ref{Bodl}) is integrable if and only if $f(k)=1$ or
$f(k)=\coth(\frac{k}{c_1})$.  \end{Pro}

In the standard ($x$) representation  equation (\ref{Bodl}) has the form
\[
u_t=\frac{1}{2\pi}\int
\xi_1^2f(\xi_1)u(\xi_1)e^{i\xi_1x}d\xi_1+uu_1=T(u_2)+uu_1 \, .
\]
Function $f(k)=1$ corresponds to the Burgers equation, while
$f(k)=\coth(\frac{k}{c_1})$ corresponds to the gravity waves on the
finite depth water (see for example \cite{AS}). In the last case $f(k)$
can be made nonsingular at $k=0$ by the transformation $f(k)\to
f(k)-c_1/k$ which corresponds to the Galilean transformation $\omega(k)
\to\omega(k)-c_1k$.  The Benjamin-Ono equation corresponds to the limit
$c_1\to 0$.

{\bf Proof} Quasi-local extension by
pseudo-differential operator $T$ can be done in a similar was as withe the
Hilbert operator $\hat{H}$ (\ref{R^n}). Coefficient
$\phi_1(\xi_1,\eta )$ of the corresponding formal recursion
operator $\Lambda=\eta+u\phi_1(\xi_1,\eta
)+u^2\phi_2(\xi_1,\xi_2,\eta )+u^3\phi_3(\xi_1,\xi_2,\xi_3,\eta )
+\ldots$ is quasi-local. Indeed,
\[
\phi_1(\xi_1,\eta )=\frac{\xi_1a(\xi_1,\eta )}{\omega(\xi_1+\eta
)-\omega(\xi_1)-\omega(\eta )}= \frac{a(\xi_1,\eta
)}{2\eta(1+\frac{\xi_1-\xi_1f(\xi_1)}{2\eta})} ,\quad
\eta\to\infty
\]
Let us consider the expansion on $1/\eta$ of the next coefficient
$\phi_2(\xi_1,\xi_2,\eta )$
\[
\phi_2(\xi_1,\xi_2,\eta
)=\phi_{22}(\xi_1,\xi_2)\eta^{-2}+\phi_{23}(\xi_1,\xi_2)\eta^{-3}+
\phi_{24}(\xi_1,\xi_2)\eta^{-4}+\ldots,
\]
where functions $\phi_{22}(\xi_1,\xi_2)$ and
$\phi_{23}(\xi_1,\xi_2)$ are quasi-local (we do not present
here explicit expressions, they are rather large), while the
function $\phi_{24}(\xi_1,\xi_2)$ can be represented in the form
\[
\phi_{24}(\xi_1,\xi_2)=\frac{F(\xi_1,\xi_2)}{\xi_1+\xi_2},
\]
where $F(\xi_1,\xi_2)$ is a quasi-local function. In general, function $\phi_{24}(\xi_1,\xi_2)$ does not
corresponds to any element of our extended differential ring and
is the obstacle to the integrability. This obstacle would be absent if
$F(\xi_1,\xi_2)$ divisable by $\xi_1+\xi_2$. The
 conditions of divisibility are
\begin{equation}\label{odd}
c_1(f(-\xi_1)+f(\xi_1))=0
\end{equation}
and
\[
(f(-\xi_1)+f(\xi_1))(f(-\xi_1)f(\xi_1)-c_0f(-\xi_1)-c_0f(\xi_1)+1)=0\, .
\]
If $f(k)$
is regular at zero and is not an odd function $c_1=0$ then from
the last equation it is easy to see that $f(k)\equiv 1$. If
$f(-k)=-f(k)$ then it can be shown that the coefficient
$\phi_2(\xi_1,\xi_2,\eta )$ is quasi-local. Making the
expansion of the next coefficient $\phi_3(\xi_1,\xi_2,\xi_3,\eta
)$ we obtain $$ \phi_3(\xi_1,\xi_2,\xi_3,\eta
)=\phi_{33}(\xi_1,\xi_2,\xi_3)\eta^{-3}+\phi_{34}(\xi_1,\xi_2,\xi_3)
\eta^{-4}+\ldots , $$ where the function
$\phi_{33}(\xi_1,\xi_2,\xi_3)$ is quasi-local while
$\phi_{34}(\xi_1,\xi_2,\xi_3)$ can be represented in the form $$
\phi_{34}(\xi_1,\xi_2,\xi_3)=\frac{G(\xi_1,\xi_2,\xi_3)}{\xi_1+\xi_2+\xi_3},
$$ where $G(\xi_1,\xi_2,\xi_3)$ is a quasi-local function. Coefficient
$\phi_{34}(\xi_1,\xi_2,\xi_3)$ is is quasi-local if $G(\xi_1,\xi_2,\xi_3)$
is divasable by $\xi_1+\xi_2+\xi_3$ and the divisibility condition yealds the following differential
equation $$ -6g+6zg'+z^2g''=0,\quad g=f^2+c_1f'-1 \, , $$
which general solution is
$$
g=\alpha z+\frac{\beta}{z^6} $$ Taking into account the Laurent
series at zero we have $\alpha=\beta=0$ and (\ref{odd}) we find
\[
 f=\coth(\frac{z}{c_1}) \, .
\]
$\quad_{\triangle}$

\subsection{Camassa-Holm-Degasperis equation}

Equation (\ref{CHD}) can be obviously rewritten as scalar evolution
nonlocal equation
\begin{equation}
\label{CHD1} u_t=\Delta (-uu_3+(c+1)uu_1-cu_1u_2)\, ,\quad c\ne
0\, ,
\end{equation}
where operator $\Delta=(1-D^2)^{-1}$. It is also well known that
its higher symmetries (when they exist) contain nested operator
$\Delta$ and we need to extend the differential ring $\ring$ in a
similar as in the previous section
\begin{equation}\label{R^nCHD}
{\cal R}^0_\Delta=\ring\, , \quad {\cal
R}^1_\Delta=\overline{{\cal R}^0_\Delta\bigcup \Delta({\cal
R}^0_\Delta)} \, ,\quad {\cal R}^{n+1}_\Delta=\overline{{\cal
R}^n_\Delta\bigcup \Delta({\cal R}^n_\Delta)}\, ,
\end{equation}

Symbolic representation of operator $\Delta$ is
$\Delta\to\frac{1}{1-\eta^2}$. The symbolic representation of
elements of differential rings ${\cal R}^n_\Delta$ is obvious. For
example if $A$ is an element from ${\cal R}^0_\Delta$ with
corresponding symbol $u^n a(\xi_1,...,\xi_n)$ then $\Delta(A)$
has a symbol $u^n\frac{a(\xi_1,\ldots,\xi_n)}
{1-(\xi_1+\cdots+\xi_n)^2}$.

The procedure of testing the integrability of given equation is
exactly the same as in the previous case. Let us apply it to the
Camassa-Holm-Degasperis type equations.
\begin{The}
Equation (\ref{CHD1}) is integrable only if $c=2$ or $c=3$
\end{The}
{\bf Proof}

In order to  introduce a linear term in (\ref{CHD1}) we shift
$u\to u-1$
\[
u_t=(1-D^2)^{-1}(u_3-(c+1)u_1-uu_3+(c+1)uu_1-cu_1u_2)
\]
In the symbolic representation it can be rewritten as follows
\[
u_t=u\omega(\xi_1)+\frac{u^2}{2}a(\xi_1,\xi_2)=F
\]
where
\[
\omega(k)=\frac{k^3-(c+1)k}{1-k^2}
\]
\[
a(\xi_1,\xi_2)=\frac{(c+1)(\xi_1+\xi_2)-(\xi_1^3+\xi_2^3)-c\xi_1\xi_2(\xi_1+\xi_2)}{1-(\xi_1+\xi_2)^2}
\]
Calculating first two coefficients of corresponding formal
recursion operator $$ \Lambda=\eta+u \phi_1(\xi_1,\eta )+u^2
\phi_2(\xi_1,\xi_2,\eta )+\cdots $$. The first coefficient
\[
\phi_1(\xi_1,\eta
)=\frac{(\xi_1^2-1)(\eta^2-1)(\xi_1^2+\eta^2-\xi_1\eta-1+c(\xi_1\eta-1))}
{ c \eta (\eta^2+\xi_1^2+\xi_1\eta-3)}
\]
is quasi-local because coefficients of its expansion in $1/\eta$
are polynomials on $\xi_1$.  For the second coefficient
$\phi_2(\xi_1,\xi_2,\eta )$ we have the following expansion
\[
\phi_2(\xi_1,\xi_2,\eta
)=\Phi_{21}(\xi_1,\xi_2)\eta+\Phi_{20}(\xi_1,\xi_2)+\Phi_{2,-1}
(\xi_1,\xi_2)\eta^{-1}+
\]
\[
+\Phi_{2,-2}(\xi_1,\xi_2)\eta^{-2}+\Phi_{2,-3}(\xi_1,\xi_2)\eta^{-3}+\cdots,
\]
where coefficients $\Phi_{21}(\xi_1,\xi_2),\ldots,
\Phi_{2,-2}(\xi_1,\xi_2)$ are polynomials on its arguments (we do
not present here explicit expressions for this functions - they
are quite large), while the coefficient $\Phi_{2,-3}$ has the form
\[
\Phi_{2,-3}(\xi_1,\xi_2)=\frac{f(\xi_1,\xi_2)}{1-\xi_1\xi_2}
\]
and $f(\xi_1,\xi_2)$ is polynomial. If $f(\xi_1,\xi_2)$ does not
have $1-\xi_1\xi_2$ as a factor, then the symbol $u^2\Phi_{2,-3}$
does not correspond any element of our extended ring and hence is
the obstacle to integrability for equation (\ref{CHD1}).
Polynomial $f(\xi_1,\xi_2)$ can be divided by $1-\xi_1\xi_2$ only
if the condition
\[
(c-2)(c-3)=0
\]
is satisfied and in these cases the coefficient
$\Phi_{2,-3}(\xi_1,\xi_2)$ is  a polynomial. It is well known
that equation (\ref{CHD1}) is integrable if $c=2,3$ (see
\cite{CH,D,DHH}). $\quad_{\triangle}$

\section*{Acknowledgments}
We would like to thank A. Hone, A.B. Shabat, V.V. Sokolov, P. Santini and Jing Ping Wang
for useful discussions. This work was partially done during the six month programme ``Integrable Systems'' at the Isaac Newton Institute, Cambridge.
It was partially supported by the Royal Society (Leeds-Landau join research project). VSN thanks P. Lushnikov, D. Holm and I. Gabitov from the Los Alamos National Laboratory  where he was temporary
employed as Staff Research Ass. (CNLS and T-7 groups, Theoretical
division), INTAS grant \# 1789 and University of Julich (Landau) Scolarship.

\end{document}